\documentstyle[12pt]{article}

\def\presentation{
\voffset -.50in
\hoffset -.19in
\oddsidemargin 0in \evensidemargin 0in
\marginparwidth .75in \marginparsep 7pt \topmargin 0in
\headheight 12pt \headsep .25in
\footheight 18pt \footskip .35in
\textheight 9.5in \textwidth 6.5in
\columnsep 10pt \columnseprule 0pt }

\presentation

\newcommand{\beq}{\begin{equation}}
\newcommand{\eeq}{\end{equation}}
\newcommand{\bea}{\begin{eqnarray}}
\newcommand{\bean}{\begin{eqnarray*}}
\newcommand{\eea}{\end{eqnarray}}
\newcommand{\eean}{\end{eqnarray*}}

\newcommand{\sumi}{\sum_{i=1}^N}
\newcommand{\sumij}{\sum_{_{\ i \neq j}^{i,j=1}}^N}
\newcommand{\la}{\lambda}
\newcommand{\si}{\sigma}
\newcommand{\tr}{\mbox{tr}}
\newcommand{\Tr}{\mbox{Tr}}
\newcommand{\e}{\mbox{e}}
\newcommand{\D}{{\cal D}}
\newcommand{\q}{q}
\newcommand{\CC}{\mbox{C}}

\newtheorem{th}{Theorem}[section]

\newtheorem{lem}{Lemma}[section]

\def\C{\bf{C}}
\def\blsquare{\hfill
{\vrule height6pt width6pt depth1pt} \break \vspace{.01cm}}

\title{The Gervais-Neveu-Felder equation and the quantum Calogero-Moser
systems}
\author{J. Avan $^*$ \and O. Babelon $^*$ \and E. Billey
\thanks{L.P.T.H.E. Universit\'e Paris VI (CNRS UA 280), Box 126, Tour 16,
$1^{{\rm er}}$ \'etage, 4 place Jussieu, 75252 Paris Cedex 05, France}}
\date{May 1995}

\begin{document}

\begin{titlepage}
\renewcommand{\thepage}{}
\maketitle
\vspace{2cm}
\begin{abstract}
We quantize the spin Calogero-Moser model in the $R$-matrix formalism. The
quantum $R$-matrix of the model is dynamical. This $R$-matrix has already
appeared in Gervais-Neveu's quantization of Toda field theory and in Felder's
quantization of the Knizhnik-Zamolodchikov-Bernard equation.
\end{abstract}

\vfill

PAR LPTHE 95-25

\end{titlepage}
\renewcommand{\thepage}{\arabic{page}}

\section{Introduction}

Integrable systems of $N$ particles on a line with pairwise interaction
have recently attracted much attention. After the famous works of
Calogero and Moser \cite{CalMo}, many generalizations have been
proposed. These include the relativistic generalization of
Ruijsenaars \cite{Rui}, the spin generalization of the non--relativistic
models \cite{GibHe,BeGauHaPa} and finally the spin generalization of the
relativistic models \cite{KriZa}. They have many relations to harmonic
analysis \cite{OlPer}, algebraic geometry \cite{HiNe},
topological field theory \cite{GorNe}, conformal field theory
\cite{Fel1,EtKir}, string field theory \cite{AvJev}.

In this paper we consider yet another aspect of these models, i.e. their
embedding into the $R$-matrix formalism, both at the
classical and quantum levels.
In this respect the essentially new feature which emerges is that the
$R$-matrix turns out to be a dynamical one. At the classical level, the
$r$-matrix was computed for the usual Calogero-Moser models in
\cite{AvTaSkBraSu}.
It was computed in \cite{BiAvBa} for their spin generalization, while
it was calculated first in the sine-Gordon soliton case \cite{BaBe}
then in the general case \cite{AvRol} for the
Ruijsenaars systems.
We address here the issue of the quantum formulation of these models within
an $R$-matrix framework. We are going to show that the quantum Yang-Baxter
equation has to be generalized. At present this new equation stands at
the crossroads of three seemingly distinct topics: quantization of
Toda field theory, quantization of KZB equations, and quantization of
Calogero-Moser-Ruijsenaars models.

In section 2 we explain the above connections at the classical level.
The classical $r$-matrix of the Calogero-Moser model, the KZB connection
for the WZW model on the torus and the $r$-matrix of the exchange
algebra in Toda field theory all satisfy the same generalized Yang-Baxter
equation.
In section 3 we take advantage of these identifications to define the
commutation relations obeyed by the quantum Lax operator of the
Calogero-Moser model.
In section 4 we use this quantum algebra to construct a set of commuting
operators which are the quantum analogs of $\tr L^n$ where $L$ is the Lax
matrix of the
system.
Finally in section 5 we give examples of such operators built for
specific representations of the quantum algebra.

One should stress again that our concern here was to embed the Calogero-Moser
systems into
the $R$-matrix formalism. Many other different approaches exist for these
models. In
particular the works \cite{BeGauHaPa,EtKir} are probably closely
 related to our results, but
the precise connexions are yet to be clarified.

\section{The generalized classical Yang-Baxter equation}

\subsection{The Calogero-Moser model and its classical $r$-matrix}

The Calogero-Moser system is a system of $N$ particles on a line with
positions $x_i$ and momenta $p_i.$ The Hamiltonian is:
\beq
\label{Hstandard}
H = \frac{1}{2} \sumi p_i^2 - \frac{1}{2} g^2 \sumij V(x_{ij}), \quad
x_{ij}=x_i-x_j
\eeq
where the two-body potential $V(x)$ is the Weierstrass function $\wp(x)$
or its trigonometric limit $1/\sinh^2(x)$, or its rational limit $1/x^2.$
The Poisson bracket is the canonical one:
$$ \{p_i,x_j\}=\delta_{ij}. $$
Rather than considering the Calogero-Moser model in this standard version,
it will be important to consider instead its spin generalization
\beq
H = \frac{1}{2} \sumi p_i^2 - \frac{1}{2} \sumij h_{ij} h_{ji} V(x_{ij})
\eeq
where the Poisson bracket on the new dynamical variables $h_{ij}$ is
given by
$$ \{h_{ij},h_{kl}\}=\delta_{li} h_{kj} - \delta_{jk} h_{il}. $$
The above Poisson bracket is degenerated. We have to choose
particular symplectic leaves which we parametrize as
$$ h_{ij} = \sum_{\alpha=1}^l b_i^{\alpha} a_j^{\alpha} $$
with $$ \{a_i^{\alpha},b_j^{\beta}\}=-\delta_{\alpha\beta} \delta_{ij}. $$
Remark that $\{H,h_{ii}\}=0$ for all $i.$ The standard
Calogero-Moser model is then obtained by a Hamiltonian reduction of
the spin model for $l=1$ under this symmetry.
Indeed in this case we have $h_{ij} = a_i b_j$; the reduced manifold is
characterized by the value of the momentum $h_{ii} = a_i b_i = g$ for all
$i$. Then $h_{ij} h_{ji} = g^2$ and we recover eq.(\ref{Hstandard}).

The standard Calogero-Moser model is well known to be integrable. It has
a Lax matrix depending on a spectral parameter $\la$ \cite{Kri}
$$ L_{ij}(\la,x,p) = \delta_{ij} p_i + (1-\delta_{ij}) \Phi(x_{ij},\la) $$
with
$$ \Phi(x,\la)=\frac{\si(\la - x)}{\si(x) \si(\la)} $$
where $\si$ is the Weierstrass $\si$ function.

This yields conserved quantities $I_n=\tr L^n.$ However Liouville
integrability requires that these quantities be in involution. This is
equivalent \cite{BaVi} to the existence of an $r$-matrix (we use the standard
notation
$L_1 = L \otimes Id $ ...)
\beq
\label{rLax}
\{L_1,L_2\}=[r_{12},L_1]-[r_{21},L_2].
\eeq
This $r$-matrix was computed in \cite{AvTaSkBraSu} and is given by
\bea
\label{standardr}
r_{12}^{Cal}(\la,\mu,x)= &-& \sumij \Phi(x_{ij},\la-\mu) e_{ij} \otimes e_{ji}
+\zeta(\la-\mu) \sumi e_{ii} \otimes e_{ii} \nonumber \\
&+& \sumij \Phi(x_{ij},\mu) e_{ii} \otimes e_{ij}.
\eea
The important new feature of this model is that the $r$-matrix
depends on the dynamical variables $x_i.$

Occurence of the last term in eq.(\ref{standardr}) jeopardizes
the eventual quantization of eq.(\ref{rLax}). It is in this
context that the consideration of the "spin" model will be advantageous.
Defining
\beq
\label{Lspin}
L_{ij}(\la,x,p) = \delta_{ij} p_i + (1-\delta_{ij}) h_{ij}
\Phi(x_{ij},\la)
\eeq
we find
\beq
\label{rLaxtype}
\{L_1,L_2\}=[r_{12},L_1]-[r_{21},L_2]+[{\cal D},r_{12}]
\eeq
with
\beq
\label{spinr}
r_{12}(\la,\mu,x)=-\sumij \Phi(x_{ij},\la-\mu) e_{ij} \otimes e_{ji}
                  +\zeta(\la-\mu) \sumi e_{ii} \otimes e_{ii}
\eeq
and
\beq
{\cal D}=\sumi h_{ii} \frac{\partial}{\partial x_i}.
\eeq
The last term $[{\cal D},r_{12}]$ reflects the non-integrability of the
non-reduced system. Since the matrix $r$ only depends on the
differences $x_{ij}=x_i-x_j$, the last term takes the explicit form
$$[{\cal D},r_{12}(\la,\mu,x)]=\frac{1}{2} \sumij (h_{ii}-h_{jj})
\frac{\partial}{\partial x_{ij}} r_{12}(\la,\mu,x).$$
Its contributions will eventually vanish on the reduced phase space
$h_{ii}=\mbox{constant}.$ One can recover the $r$-matrix (\ref{standardr})
from (\ref{spinr}) using the reduction procedure \cite{BiAvBa}.

\proclaim Proposition.
The $r$-matrix eq.(\ref{spinr}) is antisymmetric:$r_{12}(\lambda, \mu,x ) = -
r_{21}(\mu,\lambda,x)$ and satisfies the equation
\beq
\label{Jacobi1}
-\{ L_1, r_{23} \} + \{ L_2, r_{13} \} - \{ L_3, r_{12} \}
+ [r_{12}, r_{13}] + [r_{12}, r_{23}] + [r_{13}, r_{23}] = 0.
\eeq
In particular, this implies the Jacobi identity.
\par

{\it Proof.}
Denoting
$ Z_{12} = [ {\cal D}, r_{12}]$, the Jacobi identity reads
\bea
\label{Jac}
0&=& \{ L_1, \{ L_2, L_3 \} \}+ \{ L_2, \{ L_3, L_1 \} \}  + \{ L_3, \{ L_1,
L_2 \} \}
\nonumber \\
&=&\relax [ L_1 ,[r_{12},r_{23}] + [r_{12}, r_{13}] + [r_{32},r_{13}]
+ \{ L_2, r_{13} \} - \{ L_3, r_{12} \} ] + \mbox{cycl. \ perm.} \nonumber\\
&+& [ r_{23},Z_{12}] + [r_{31}, Z_{23}] + [r_{12}, Z_{31}]
- [r_{32}, Z_{13}] - [r_{13}, Z_{21}] - [r_{21},Z_{32}] \\
&+&\{ L_1, Z_{23} \} + \{ L_2, Z_{31} \} + \{ L_3 , Z_{12} \} \nonumber
\eea
Using the antisymmetry of $r$, we find
\bean
[ r_{23},Z_{12}] + [r_{31}, Z_{23}] + [r_{12}, Z_{31}]
- [r_{32}, Z_{13}] - [r_{13}, Z_{21}] - [r_{21},Z_{32}] && \\
&& \hspace{-8cm} =- [ {\cal D}, [r_{12}, r_{13}] + [r_{12}, r_{23}]
+ [r_{13}, r_{23}] ].
\eean
Moreover, we have
\bean
\{ L_1, Z_{23} \} &=& \{ L_1, [ {\cal D}, r_{23}] \}
= [{\cal D}, \{ L_1, r_{23} \}] + [ \{ L_1, {\cal D} \}, r_{23}] \\
&=& [{\cal D}, \{ L_1, r_{23} \}] -  \sumi [ [ L_1,e_{ii}^{(1)} ]
\partial_{x_i}, r_{23} ] \\
&=& [{\cal D}, \{ L_1, r_{23} \}] -[ L_1 , \{ L_1, r_{23} \}]
\eean
so that eq.(\ref{Jac}) becomes
\bean
0&=&\relax [ L_1, [r_{12}, r_{13}] + [r_{12}, r_{23}] + [r_{13}, r_{23}]
-\{ L_1, r_{23} \} + \{ L_2, r_{13} \} - \{ L_3, r_{12} \} ]
+ \mbox{cycl. \ perm.}  \\
&-& [ {\cal D}, [r_{12}, r_{13}] + [r_{12}, r_{23}] + [r_{13}, r_{23}]
-\{ L_1, r_{23} \} + \{ L_2, r_{13} \} - \{ L_3, r_{12} \} ]
\eean
Hence the Jacobi identity is satisfied if eq.(\ref{Jacobi1}) holds, which is
easily checked by a direct calculation.\blsquare

{}From the facts that $\{L_1,r_{23}\}=\sumi e_{ii}^{(1)} \otimes \partial_{x_i}
r_{23}$ and $r$ depends only on the
differences $x_i-x_j$, we can rewrite eq.(\ref{Jacobi1}) as
\beq
\label{Jacobi2}
[r_{12},r_{13}]+[r_{12},r_{23}]+[r_{13},r_{23}]
-\sum_{\nu} h_{\nu}^{(1)} \frac{\partial}{\partial x_{\nu}} r_{23}
+\sum_{\nu} h_{\nu}^{(2)} \frac{\partial}{\partial x_{\nu}} r_{13}
-\sum_{\nu} h_{\nu}^{(3)} \frac{\partial}{\partial x_{\nu}} r_{12}=0
\eeq
where $\{h_{\nu}\}$ is an orthonormal basis of the Cartan subalgebra of
diagonal matrices
of $\mbox{sl}_N$ and $x=\sum_{\nu} x_{\nu} h_{\nu}.$

Let us comment on the trigonometric limit of the classical $r$-matrix
$r_{12}(\la,\mu,x)$ defined in eq.(\ref{spinr}). We remark that if
$r_{12}(\la,\mu,x)$ is a solution of eq.(\ref{Jacobi2}) such that
$\forall~\nu ,~[ h_\nu \otimes 1 + 1 \otimes h_\nu , r_{12}(\la,\mu,x) ] =0$,
 then
$$ \widetilde{r}_{12}(\la,\mu,x) = \e^{\alpha(\la) x} \otimes
\e^{\alpha(\mu) x} \ r_{12}(\la,\mu,x) \ \e^{-\alpha(\la) x} \otimes
\e^{-\alpha(\mu) x} - ( \alpha(\la) - \alpha(\mu) ) \sum_{\nu} h_{\nu}
\otimes  h_{\nu} $$
is also a solution of eq.(\ref{Jacobi2}) for any function $\alpha(\la)$.
Using this freedom we see that the
trigonometric limit of eq.(\ref{spinr}) may be recast into the form
\beq
\label{rtrigo}
r_{12}(\la,\mu,x) = \coth(\la-\mu) \sum_{i,j=1}^N e_{ij} \otimes
e_{ji} - \sumij \coth(x_{ij}) e_{ij} \otimes e_{ji}.
\eeq
Eq.(\ref{Jacobi2}) will be the cornerstone of our quantization procedure
of the Calogero-Moser model. It also appeared in two other
contexts which we now briefly recall.

\subsection{Relation to the Knizhnik-Zamolodchikov-Bernard equation}

It is well known that there is a relation between conformal field theories
and the classical Yang-Baxter equation through the Knizhnik-Zamolodchikov
equation \cite{KnZaBerFelWi}. Let $r_{ij}(z)=-r_{ji}(-z)$ be a skew-symmetric
solution of the classical Yang-Baxter equation taking values in the
tensor product ${\cal G}^{(i)} \otimes {\cal G}^{(j)}$ where ${\cal G}$ is a
simple Lie algebra. Let ${\cal H}$ be a Cartan subalgebra of ${\cal G}$.
Then the KZ connexion
$$ \nabla_i=\partial_{z_i}-\sum_{_{j \neq i}^{j=1}}^N r_{ij}(z_i-z_j) $$
has zero curvature. Hence the system of equations
\beq
\label{KZ}
\partial_{z_i} u=\sum_{_{j \neq i}^{j=1}}^N r_{ij}(z_i-z_j) u
\eeq
for a function $u(z_1,\ldots,z_N)$ on $\C^N-\cup_{i<j}\{z;z_i=z_j\}$ with
values in $V \otimes \cdots \otimes V$, where $V$ is a representation
space for ${\cal G}$, has a solution.
Eq.(\ref{KZ}) characterizes conformal blocks of the Wess-Zumino-Witten model
on the sphere. On a higher genus  Riemann surface the corresponding equations
are the  Knizhnik-Zamolodchikov-Bernard equations (in our case of interest
$g=1$); they are equations for functions
$u(z_1,\ldots,z_N,x)$ taking values in the weight zero subspace of
a tensor product of irreducible finite dimensional representations
of a simple Lie algebra ${\cal G}$ i.e.
\begin{eqnarray*}
 \forall~\nu,~~\left( \sum_{i=1}^N h^{(i)}_{\nu} \right) u =0
\end{eqnarray*}
(Here and in the following the superscript in $ h^{(i)}_{\nu}$ denotes the
space
on which $h_\nu$ acts and the subscript $\nu$ denotes an element in a basis of
${\cal
H}$.)
In the case of a torus, they take
the form
\beq
\label{KZB}
\kappa \partial_{z_i} u=-\sum_{\nu} h_{\nu}^{(i)} \partial_{x_{\nu}} u
       + \sum_{_{j \neq i}^{j=1}}^N r_{ij}(z_i-z_j,x) u
\eeq
with additional equations involving derivatives w.r.t. the modular
parameters. The compatibility condition of eq.(\ref{KZB}) is exactly
eq.(\ref{Jacobi2}) \cite{Fel1}.

\subsection{Relation to Toda field theory}

The Toda field equations associated to a simple Lie algebra ${\cal G}$ read
$$ \Box \phi_{\nu}=\sum_{\alpha ~~{simple}} \alpha_{\nu} \
\mbox{e}^{2 \alpha(\phi)} $$
where $\phi=\sum_{\nu} \phi_{\nu} h_{\nu}$ is a field taking values
in a Cartan subalgebra ${\cal H}$ of a Lie algebra ${\cal G}$. As above, $ \{
h_{\nu} \}$
is an orthonormal basis of this Cartan subalgebra. In the case
${\cal G}=\mbox{sl}_2$, this becomes the Liouville equation.

Leznov and Saveliev \cite{LezSav} found a generalization of Liouville's
solution to the Liouville equation. It takes the form
$$ \mbox{e}^{- \Lambda(\phi)}=\Psi(z) \cdot \overline{\Psi}(\bar{z}) $$
where $|\Lambda \rangle$ is a highest weight vector,
$\Psi(z)$ and $\overline{\Psi}(\overline{z})$ are chiral fields ($z=\sigma +
\tau$ and $\bar{z}=
\sigma -\tau $ are the light-cone coordinates)
$$ \Psi = \langle\Lambda|Q_+, \quad \overline{\Psi} = Q_- |\Lambda \rangle $$
with $Q_{\pm}$ solutions of the linear systems
$$ \partial_z Q_+ = (P+{\cal E}_+) Q_+, \quad \partial_{\bar{z}} Q_- =
Q_- (\overline{P}+{\cal E}_-) $$
$P$ and $\overline{P}$ are
chiral
fields with values in a Cartan subalgebra of ${\cal G}$ and $ {\cal E}_{\pm} =
\sum_{\alpha ~~{simple}} E_{\pm \alpha} $ with $ E_{\alpha} $
the root vectors in the corresponding Cartan decomposition of ${\cal G}.$

To reconstruct periodic solutions of the Toda field equation, it is natural
to consider the quasi-periodic basis for $\Psi$ and $\overline{\Psi}$
$$ \Psi(\sigma+2 \pi)=\Psi(\sigma) \exp(x), \quad
\overline{\Psi}(\bar{\sigma}+2 \pi)=
\exp(-x) \overline{\Psi}(\bar{\sigma}) $$
where $x=\sum_{\nu} x_{\nu} h_{\nu}$ is the quasi momentum (zero mode),
belonging to the Cartan subalgebra.

The Poisson bracket (at equal time $\tau =0$)
$$ \{P(\sigma),P(\sigma')\} = \delta'(\sigma-\sigma') \sum_{\nu} h_{\nu}
\otimes h_{\nu}$$
induces a Poisson bracket on $\Psi$ \cite{GeNe}
\beq
\label{Poissonpsi}
\{\Psi_1(\sigma),\Psi_2(\sigma')\}=\Psi_1(\sigma) \Psi_2(\sigma') \,
r_{12}^{\pm}(x), \quad
\pm=\mbox{sign}(\sigma-\sigma')
\eeq
where \cite{BaBoToBaDaFe} (in the $\mbox{sl}_N$ case)
\beq
\label{rToda}
r_{12}^{\pm}(x)=\pm \sum_{i,j=1}^N e_{ij} \otimes e_{ji}
- \sumij \coth(x_{ij}) e_{ij} \otimes e_{ji}.
\eeq
Taking into account that
$$ \{\Psi(\sigma),x_{\nu}\}=\Psi(\sigma) h_{\nu}, $$
the Jacobi identity
$\{\Psi_1,\{\Psi_2,\Psi_3\}\}+\mbox{cycl. \ perm.}=0$ implies
exactly eq.(\ref{Jacobi2}) on $r^{\pm}(x).$

The solution (\ref{rToda}) of eq.(\ref{Jacobi2}) is related to
the solution (\ref{rtrigo}) by the formula:
$$ r_{12}(\la,x)=\frac{\e^{\la} \, r_{12}^+(x)-\e^{-\la} \, r_{12}^-(x)}
{\e^{\la}-\e^{-\la}}. $$

\section{The Gervais-Neveu-Felder equation}

In this section, we give the quantum version of eq.(\ref{Jacobi2}). This
results into a deformed version of the Quantum Yang-Baxter equation,
which first appeared in \cite{GeNe} and later in\cite{Fel1}.

We need to introduce some notations. If ${\cal G}$ is a simple Lie algebra
and ${\cal H}$ a Cartan subalgebra of ${\cal G}$, let $x=\sum_{\nu}
x_{\nu} h_{\nu}$ be an element of ${\cal H}$.
For any function $f(x)=f(\{x_{\nu}\})$ with values in $\C$, we denote
$$ f(x+\gamma h^{(i)})=\e^{\gamma \D^{(i)}} f(x) \e^{-\gamma \D^{(i)}} $$
where
$$\D^{(i)}=\sum_{\nu} h_{\nu}^{(i)} \partial_{x_{\nu}}.$$

Suppose $V^{(1)}, \ldots, V^{(n)}$ are finite dimensional diagonalizable
${\cal H}$-modules; the Gervais-Neveu-Felder equation is an equation for a
function $R_{ij}(\la,x)$ meromorphic in the spectral parameter $\la$,
depending on $x$, and taking values in $\mbox{End} (V^{(i)} \otimes
V^{(j)}).$ It reads
\bea
\label{GNF}
&& R_{12}(\la_{12},x+\gamma h^{(3)}) R_{13}(\la_{13},x-\gamma h^{(2)})
R_{23}(\la_{23},x+\gamma h^{(1)}) \nonumber \\
&& \hspace{3cm} = R_{23}(\la_{23},x-\gamma h^{(1)})
R_{13}(\la_{13},x+\gamma h^{(2)}) R_{12}(\la_{12},x-\gamma h^{(3)}).
\eea
We have used the notation $\la_{ij}=\la_i-\la_j.$

The classical limit of the Gervais-Neveu-Felder equation is obtained as usual
by expanding $R$ in powers of $\hbar=-2 \gamma$
$$ R_{12}(\la,x)=\mbox{Id}-2 \gamma \, r_{12}(\la,x)+\mbox{O}(\gamma^2). $$
The first non trivial term of eq.(\ref{GNF}) is of order $\gamma^2$ and
stems from:
\begin{eqnarray*}
\gamma ( r_{12}(x-\gamma h^{(3)})-r_{12}(x+\gamma h^{(3)}) &+& \\
 r_{13}(x+\gamma h^{(2)})-r_{13}(x-\gamma h^{(2)})  &+& \\
 r_{23}(x-\gamma h^{(1)})-r_{23}(x+\gamma h^{(1)} )) &+&  \\
 \gamma^2 ( [r_{12},r_{13}]+[r_{12},r_{23}]
+[r_{13},r_{23}] )
&+& \mbox{O}(\gamma^3) = 0.
\end{eqnarray*}
The term of order $\gamma^2$ yields exactly eq.(\ref{Jacobi2}).

Gervais and Neveu first obtained eq.(\ref{GNF}) in \cite{GeNe} as a result of
the
quantization of Liouville field theory ( this result was later extended to
$\mbox{sl}_N$ Toda
field theory in \cite{BiGe}). In this quantization procedure, the
quantum version of eq.(\ref{Poissonpsi}) was shown to take the form
of an exchange algebra:
\beq
\label{exchange}
\Psi_1(\sigma) \Psi_2(\sigma') = \Psi_2(\sigma') \Psi_1(\sigma)
R_{GN}^{\pm}(x,q), \quad
\pm = \mbox{sign}(\sigma-\sigma'), \quad q=\e^{-2 \gamma},
\eeq
where for $\mbox{sl}_N$ \cite{BiGe,BoBo}
\bea
\label{RGervaisNeveu}
R_{GN}^{\pm}(x,q)= q^{\mp 1/N} \left [
q^{\pm 1} \sumi e_{ii} \otimes e_{ii} \right.
&+& \sumij \frac{q \e^{x_{ij}} - q^{-1} \e^{-x_{ij}}}
{\e^{x_{ij}} - \e^{-x_{ij}}} e_{ii} \otimes e_{jj} \nonumber \\
&-& (q-q^{-1}) \sumij \left. \frac{\e^{\mp x_{ij}}}
{\e^{x_{ij}} - \e^{-x_{ij}}} e_{ij} \otimes e_{ji} \right ].
\eea
Taking into account the shift property of the fields $\Psi$, that is,
for any scalar function $f(x)$:
\beq
\label{shift}
f(x) \ \Psi_1(\sigma) = \Psi_1(\sigma) \ f(x-2 \gamma h^{(1)}), \quad
q=e^{-2\gamma},
\eeq
the associativity of the $\Psi$ fields algebra yields
\begin{eqnarray}
R_{12}(x)R_{13}(x-2 \gamma h^{(2)}) R_{23}(x)=
R_{23}(x-2 \gamma h^{(1)}) R_{13}(x) R_{12}(x-2 \gamma h^{(3)})
\nonumber
\end{eqnarray}
This is equivalent to eq.(\ref{GNF}) if $R_{ij}(x)$ satisfies the relation
\begin{eqnarray}
\relax [ {\cal D}^{(i)}+ {\cal D}^{(j)}, R_{ij}(x) ] =0
\label{20bis}
\end{eqnarray}
which is true for the $R$-matrix given by eq.(\ref{RGervaisNeveu}).

Felder \cite{Fel1} interpreted eq.(\ref{GNF}) as a compatibility
condition for the algebra of  $L$-operators (following the well known
Leningrad school approach \cite{Fadd}):
\bea
\label{RLL}
R_{12}(\la_{12},x+\gamma h^{(\q)}) \widetilde{L}_{1 \q}(\la_1,x-\gamma h^{(2)})
\widetilde{L}_{2 \q}(\la_2,x+\gamma h^{(1)}) && \nonumber \\
&& \hspace{-8cm}= \widetilde{L}_{2 \q}(\la_2,x-\gamma h^{(1)})
\widetilde{L}_{1 \q}(\la_1,x+\gamma h^{(2)}) R_{12}(\la_{12},x-\gamma
h^{(\q)}).
\eea
Here we assume that the matrix elements of
$\widetilde{L}_{1 \q}$ act on a quantum space $V^{(q)}$ which is
a ${\cal H}$-module so that the action of $h^{(q)}$ is defined.
In the following we will be interested in $R$-matrices and representations
$ \widetilde{L}_{i \q} $ satisfying the properties
\bea
\label{comR}
[ h^{(i)}+h^{(j)} , R_{ij}(\la_{ij},x) ] &=& 0 \\
\label{hL}
[ h^{(i)} + h^{(\q)} , \widetilde{L}_{i \q} ] &=& 0.
\eea
{ From }eq.(\ref{shift}) we see that $\Psi(z)$ in the exchange algebra
(\ref{exchange}) naturally contains the
shift operator $\e^{2\gamma\D}$. By analogy we define a Lax operator:
\beq
\label{newL}
L_{i \q}(\la,x) = \e^{\gamma \D^{(i)}} \widetilde{L}_{i \q}(\la,x)
\e^{\gamma \D^{(i)}}.
\eeq
In the limit when $\gamma \longrightarrow 0$, and assuming that
$ \widetilde{L}(\la,x) = \mbox{Id} + 2 \gamma \widetilde{l}(\la,x)
+ \mbox{O}(\gamma^2)$, the behaviour of $L$ is
$$ L(\la,x) = \mbox{Id} + 2 \gamma \left ( \sum_{\nu}
h_{\nu} \frac{\partial}{\partial x_{\nu}} + \widetilde{l}(\la,x) \right )
+ \mbox{O}(\gamma^2), $$
which is the typical form (see eq.(\ref{Lspin})) of the Lax matrix of the
Calogero-Moser system. The shift operator $\e^{\gamma \D}$ thus
contributes to reintroducing the momentum $p_{\nu}=\partial_{x_{\nu}}$
on the diagonal.

This operator (\ref{newL}) now obeys the following equation:
\beq
\label{newRLL}
R_{12}(\la_{12},x+\gamma h^{(\q)}) L_{1 \q}(\la_1,x) L_{2 \q}(\la_2,x)
= L_{2 \q}(\la_2,x) L_{1 \q}(\la_1,x) R_{12}(\la_{12},x-\gamma h^{(\q)})
\eeq
provided one has
\beq
\label{comDR}
[ \D^{(1)}+\D^{(2)}, R_{12}(\la_{12},x) ] = 0.
\eeq
Eq.(\ref{hL}) translates into the following properties:
\bea
\label{shift1}
f(x-\gamma h^{(\q)}) \, L_{i \q} &=& L_{i \q} \, f(x-\gamma h^{(\q)}-2
\gamma h^{(i)}) \\
\label{shift2}
f(x+\gamma h^{(\q)} + 2 \gamma h^{(i)}) \, L_{i \q} &=& L_{i \q} \,
f(x+\gamma h^{(\q)}).
\eea
As in the classical case, if $R_{12}(\la_{12},x)$ is a solution of
eq.(\ref{GNF}) having the property $ [ h^{(1)}+h^{(2)} , R_{12}(\la_{12},x)
] = 0 $, then
\bea
\label{qtransf}
\widetilde{R}_{12}(\la_{12},x) = \e^{[\alpha(\la_1)+\beta]x} \otimes
\e^{[\alpha(\la_2)-\beta]x} \ \e^{\gamma [\alpha(\la_1)-\alpha(\la_2)-\beta]
h \otimes h} \ R_{12}(\la_{12},x) && \nonumber \\
&& \hspace{-6cm} \e^{\gamma [\alpha(\la_1)-\alpha(\la_2)+\beta]
h \otimes h} \ \e^{-[\alpha(\la_1)-\beta]x}  \otimes
\e^{-[\alpha(\la_2)+\beta]x},
\eea
 defines another solution of eq.(\ref{GNF})
with $\alpha(\la)$  an arbitrary function of $\la$ and $\beta$  an
arbitrary parameter.
A solution of eq.(\ref{GNF}), the classical limit of which -- up to a
redefinition of type (\ref{qtransf}) -- is the
$r$-matrix (\ref{spinr}), was given in \cite{Fel1}. It reads
\beq
\label{RFelder}
R_{F}(\la,x)= \sumi e_{ii} \otimes e_{ii}
+ \sumij \frac{\si(\la) \si(2 \gamma-x_{ij})}{\si(2 \gamma-\la) \si(x_{ij})}
e_{ii} \otimes e_{jj} + \sumij \frac{\si(2 \gamma) \si(x_{ij}-\la)}
{\si(2 \gamma-\la) \si(x_{ij})}
e_{ij} \otimes e_{ji}.
\eeq
Just as in the classical case, the relation between (\ref{RFelder})
and (\ref{RGervaisNeveu}) is obtained by using transformation (\ref{qtransf})
and by taking the trigonometric limit.
One gets
\beq
\label{RF-RGN}
R_{F}(\la,x) = \frac{\e^{\la} q^{1/N} R_{GN}^+(x,q)
- \e^{-\la} q^{-1/N} R_{GN}^-(x,q)}{q \e^{\la} -
q^{-1} \e^{-\la}}, \quad q=\e^{-2 \gamma}.
\eeq
Let us recall at this point some known facts about the matrices
$R_{GN}^{\pm}(x,q).$ These matrices are related to Drinfeld's
matrices $R_{D}^{\pm}$ by:
$$ R_{GN}^{\pm}(x,q) = F_{21}^{-1}(x) \, R_{D}^{\pm} \,
F_{12}(x) $$
where $R_{D,12}^- = (R_{D,21}^+) ^{-1},$
\beq
R_{D}^+ = \sumij e_{ii} \otimes e_{jj} + q \sumi e_{ii} \otimes
e_{ii} + (q-q^{-1}) \sum_{_{\ i < j}^{i,j=1}}^N  e_{ij} \otimes e_{ji}
\eeq
and
\bean
F_{12}(x) = \sumi e_{ii} \otimes e_{ii}
&+& \sum_{_{\ i < j}^{i,j=1}}^N \frac{1}{\e^{x_{ij}} - \e^{-x_{ij}}}
e_{ii} \otimes e_{jj} + \sum_{_{\ i > j}^{i,j=1}}^N
\frac{1}{q \e^{-x_{ij}} - q^{-1} \e^{x_{ij}}} e_{ii} \otimes e_{jj} \\
&-&(q-q^{-1}) \sum_{_{\ i < j}^{i,j=1}}^N \frac{\e^{x_{ij}}}
{(\e^{x_{ij}} - \e^{-x_{ij}}) (q \e^{x_{ij}} - q^{-1} \e^{-x_{ij}})}
e_{ij} \otimes e_{ji}.
\eean
In the $\mbox{sl}_2$ case, a universal formula for $F_{12}(x)$ is
available \cite{Ba1}.

In the framework of Toda field theory, it is known that one can eliminate
the $x$ dependence from the exchange algebra (\ref{exchange}) by a suitable
change of basis \cite{Ba2,CrGer}: defining
$$ \xi(\sigma) = \Psi(\sigma) M(x) $$
with $$M(x)=\sum_{i,j=1}^N e^{2 j (x_i-\frac{1}{N} \sum x_k )}
e_{ij}$$
we get
$$ \xi_1(\sigma) \xi_2(\sigma') = \xi_2(\sigma') \xi_1(\sigma) R_{CG}^{\pm}(q),
\quad \pm=\mbox{sign}(\sigma-\sigma') $$
with an $R$-matrix $R_{CG}^{\pm}(q)$ independent of $x$ \cite{CrGer}:
\bean
R_{CG}^+(q) &=& q^{-1/N} \left ( q \sumi e_{ii} \otimes e_{ii}
+ q \sum_{_{\ i > j}^{i,j=1}}^N q^{-2 (i-j)/N} e_{ii} \otimes e_{jj}
+ q^{-1} \sum_{_{\ i < j}^{i,j=1}}^N q^{-2 (i-j)/N} e_{ii} \otimes e_{jj}
\right. \\
&-& \left. (q-q^{-1}) \sum_{_{\ i < j}^{i,j=1}}^N \sum_{r=1}^{j-i-1}
q^{2 r/N} e_{j-r,i} \otimes e_{i+r,j}
+ (q-q^{-1}) \sum_{_{\ i > j}^{i,j=1}}^N \sum_{r=0}^{i-j-1} q^{-2 r/N}
e_{j+r,i} \otimes e_{i-r,j} \right )
\eean
and of course $ R_{CG,12}^-(q) = [ R_{CG,21}^+(q) ]^{-1}. $

We may wonder whether the $x$ dependence in (\ref{newRLL}) may not be
eliminated by a similar change of variables. Starting from eq.(\ref{newRLL})
with the $R$-matrix given by (\ref{RF-RGN}), we set
$$ {\cal L}(\la,x) = M^{-1}(x+\gamma h^{(\q)}) L(\la,x) M(x-\gamma h^{(\q)}).
$$
Then eq. (\ref{newRLL}) becomes
$$ R^{\star}_{CG}(\la-\mu,q) {\cal L}_1(\la,x) {\cal L}_2(\mu,x) =
{\cal L}_2(\mu,x) {\cal L}_1(\la,x) R_{CG}(\la-\mu,q) $$
with
$$ R_{CG,12}(\la,q) = \frac{\e^{\la} R^+_{CG}(q)-
\e^{-\la} R^-_{CG}(q)}{q \e^{\la} - q^{-1} \e^{-\la} }, \quad
R^{\star}_{CG,12}(\la,q) = R_{CG,21}(-\la,q^{-1}).
$$
This equation is reminiscent of the equation studied in \cite{FoIoJiKeMiYa}.

\section{Construction of commuting operators}

We now present a set of commuting operators quantizing the classical
quantities $\tr L^n$. We consider in this section an abstract algebraic
setting.
Examples will be provided in the next section when there is no spectral
parameter
$(\lambda = \infty) $. In that case one can restrict oneself to finite
dimensional quantum groups.
 In the case with spectral parameter, one should consider full affine quantum
groups,
 and this will be left for further investigations.

In the context of the non-shifted Yang-Baxter
equation $R_{12} L_1 L_2 = L_2 L_1 R_{12}$, the quantum analogs of
the conserved quantities $\tr L^n$ are to be defined \cite{Mail} as
$$ I_n=\Tr_{1 \ldots n} \ [ L_1 \ldots  L_n \hat{R}_{12} \hat{R}_{23} \ldots
\hat{R}_{n-1,n} ] $$
where
$$ \hat{R}_{ij} = P_{ij} R_{ij} $$
and $P_{ij}$ are the permutation operators of the spaces $i$ and $j$.

In the Gervais-Neveu-Felder case, we have the following
\begin{th}
\label{theo}
Let $R(x)$ and $L$ be as in eqs.(\ref{GNF},\ref{newRLL}) with the shift
properties
as in eqs.(\ref{shift1},\ref{shift2}), and condition (\ref{20bis}) be
satisfied.

We define the operators
\beq
I_n = \Tr_{1 \ldots n} \ [ L_1(x) \ldots  L_n(x) \hat{R}_{12}(x-2 \gamma
h^{(3,n)}) \ldots \hat{R}_{k,k+1}(x-2 \gamma h^{(k+2,n)}) \ldots
\hat{R}_{n-1,n}(x) ]
\eeq
where
$$ h^{(k,l)} = \sum_{i=k}^l h^{(i)}.$$
Then:
\begin{description}
\item[1)] The operators $I_n$ leave the subspace
of zero weight vectors invariant (vectors $|V\rangle$ such that $h^{(\q)}|V
\rangle=0 $).
\item[2)] The restrictions of the operators $I_n$ to the zero weight subspaces
form a set of commuting quantities.
\end{description}
\end{th}

\noindent{\it Proof.}
To prove (1) we have to prove that $[I_n,h_{\nu}^{(\q)}]=0.$ This follows
immediately from the relations
\bean
[ h^{(\q)} , L_i ] &=& - [ h^{(i)} , L_i ] \\
\relax [ h^{(\q)} , \hat{R}_{ij} ] &=& 0 \\
\relax [ h^{(i)}+h^{(j)} , \hat{R}_{ij} ] &=& 0.
\eean
We will decompose the proof of (2) into several lemmas. We will need the
important shift properties of $L$ given by eq.(\ref{shift1},\ref{shift2}).

\begin{lem}
\label{lem1}
On the zero weight subspace, one can write
$$ I_n I_m = \Tr \ [ L_1(x) \ldots  L_n(x) L_{n+1}(x) \ldots  L_{n+m}(x)
 {\cal I}^{(1,n)}(x-2 \gamma h^{(n+1,n+m)}) {\cal I}^{(n+1,n+m)}(x) ] $$
where ${\cal I}^{(i,j)}(x) = \hat{R}_{i,i+1}(x-2 \gamma
h^{(i+2,n)}) \ldots \hat{R}_{k,k+1}(x-2 \gamma h^{(k+2,n)}) \ldots
\hat{R}_{j-1,j}(x).$
\end{lem}

\noindent{\it Proof.}
Since $I_m$ leaves the zero weight subspace invariant,
it is possible to rewrite
$$ I_n I_m = \Tr \ [ L_1(x) \ldots  L_n(x) {\cal I}^{(1,n)}(x) ] \
             \Tr \ [ L_{n+1}(x) \ldots  L_{n+m}(x) {\cal I}^{(n+1,n+m)}(x) ] $$
as
$$ I_n I_m = \Tr \ [ L_1(x) \ldots  L_n(x) {\cal I}^{(1,n)}(x-\gamma h^{(q)}) ]
          \ \Tr \ [ L_{n+1}(x) \ldots  L_{n+m}(x) {\cal I}^{(n+1,n+m)}(x) ]. $$
We now push ${\cal I}^{(1,n)}(x-\gamma h^{(q)})$ through
$L_{n+1} \ldots  L_{n+m}$ using eq.(\ref{shift1}). Applying the expression
found to the zero weight subspace gives the result.
\blsquare

\begin{lem}
\label{lem2}
We can rewrite
\begin{eqnarray*}
 I_m I_n &=& \Tr \ [ L_1(x) \ldots L_{n+m}(x)
 Q_n^{-1}(x) \ldots Q_1^{-1}(x) {\cal I}^{(n+1,n+m)}(x-2 \gamma h^{(1,n)}).
 \\
&&\hskip 9cm \cdot{\cal I}^{(1,n)}(x) Q_1(x) \ldots Q_n(x) ]
\end{eqnarray*}
where
$$ Q_i(x)=\prod_{j=1}^{^{\ m}_{\longleftarrow}} R_{i,n+j}(x-2
\gamma h^{(i+1,n)} -2 \gamma h^{(n+j+1,n+m)}). $$
\end{lem}

\noindent{\it Proof.}
According to lemma \ref{lem1},
$$ I_m I_n = \Tr \ [ L_{n+1}(x) \ldots L_{n+m}(x) L_1(x) \ldots L_n(x)
{\cal I}^{(n+1,n+m)}(x-\gamma h^{(\q)} -2 \gamma h^{(1,n)})
{\cal I}^{(1,n)}(x) ] , $$
and using
$$ L_{n+m}(x) L_1(x) = R_{1,n+m}(x+\gamma h^{(\q)}) L_1(x) L_{n+m}(x)
R_{1,n+m}^{-1}(x-\gamma h^{(\q)}) $$
we have
\bean
I_m I_n = \Tr \ [ L_{n+1}(x) \ldots L_{n+m-1}(x) R_{1,n+m}(x+\gamma h^{(\q)})
L_1(x) L_{n+m}(x)  R_{1,n+m}^{-1}(x-\gamma h^{(\q)}) && \\
&& \hspace{-10cm} L_2(x) \ldots L_n(x)
{\cal I}^{(n+1,n+m)}(x-\gamma h^{(\q)} -2 \gamma h^{(1,n)})
{\cal I}^{(1,n)}(x) ].
\eean
Using once more eq.(\ref{shift1},\ref{shift2}), eq.(\ref{comDR}) and
the cyclicity property of the trace, we get
\bean
I_m I_n = \Tr \ [ L_{n+1}(x) \ldots L_{n+m-1}(x) L_1(x) L_{n+m}(x)
L_2(x) \ldots L_n(x) R_{1,n+m}^{-1}(x-\gamma h^{(\q)} -2\gamma h^{(2,n)}) && \\
&& \hspace{-13cm} {\cal I}^{(n+1,n+m)}(x-\gamma h^{(\q)} -2 \gamma h^{(1,n)})
{\cal I}^{(1,n)}(x) R_{1,n+m}(x+\gamma h^{(\q)} -2\gamma h^{(2,n)})].
\eean
Then pushing $L_1$ through $L_{n+m-1}, L_{n+m-2}, \ldots L_{n+1}$ gives
\bean
I_m I_n = \Tr \ [ L_1(x) L_{n+1}(x) \ldots L_{n+m}(x) L_2(x) \ldots L_n(x)
Q_1^{-1}(x)  && \\
&& \hspace{-8cm} {\cal I}^{(n+1,n+m)}(x-\gamma h^{(\q)} -2
\gamma h^{(1,n)}) {\cal I}^{(1,n)}(x) Q_1(x)].
\eean
The result is obtained by repeating the procedure with $L_2, L_3, \ldots
L_n.$
\blsquare

Comparing lemmas \ref{lem1} and \ref{lem2}, commutation of $I_m$ and $I_n$
will be proved if
 \begin{eqnarray*}
 Q_1(x) \ldots Q_n(x) {\cal I}^{(1,n)}(x-2\gamma h^{(n+1,n+m)})
{\cal I}^{(n+1,n+m)}(x)  &=&  \\
&& \hskip -4cm
{\cal I}^{(n+1,n+m)}(x-2\gamma h^{(1,n)})
{\cal I}^{(1,n)}(x) Q_1(x) \ldots Q_n(x)
\end{eqnarray*}
Since ${\cal I}^{(1,n)}$ and ${\cal I}^{(n+1,n+m)}$ act on different spaces and
since $[h^{(n+1,n+m)}, {\cal I}^{(n+1,n+m)}(x)]=0,$ this last relation
is equivalent to
 \begin{eqnarray*}
 Q_1(x) \ldots Q_n(x) {\cal I}^{(n+1,n+m)}(x) {\cal I}^{(1,n)}
(x-2\gamma h^{(n+1,n+m)}) &=& \\
&&  \hskip -4cm
{\cal I}^{(n+1,n+m)}(x-2\gamma h^{(1,n)})
{\cal I}^{(1,n)}(x) Q_1(x) \ldots Q_n(x)
\end{eqnarray*}
We shall prove this relation in two steps:
\begin{description}
\item[$(\ast)$] $ Q_1(x) \ldots Q_n(x) {\cal I}^{(n+1,n+m)}(x) =
{\cal I}^{(n+1,n+m)}(x-2\gamma h^{(1,n)}) Q_1(x) \ldots Q_n(x) $
\item[$(\ast \ast)$] $ Q_n^{-1}(x) \ldots Q_1^{-1}(x) {\cal I}^{(1,n)}(x) =
{\cal I}^{(1,n)}(x-2\gamma h^{(n+1,n+m)}) Q_n^{-1}(x) \ldots Q_1^{-1}(x). $
\end{description}
Relation $(\ast)$ is a straightforward consequence of the following lemma.

\begin{lem}
\label{lem3}
Defining $T_i(x)={\cal I}^{(n+1,n+m)}(x-2\gamma h^{(i,n)}),$ we have
$$ T_i(x) Q_i(x) = Q_i(x) T_{i+1}(x). $$
\end{lem}

\noindent{\it Proof.}
Lemma \ref{lem3} will be proved if we show that
\bean
\hat{R}_{n+k,n+k+1}(x-2\gamma h^{(i,n)}-2\gamma h^{(n+k+2,n+m)}) Q_i(x) && \\
&& \hspace{-7cm} = Q_i(x) \hat{R}_{n+k,n+k+1}(x-2\gamma h^{(i+1,n)}-
2\gamma h^{(n+k+2,n+m)}).
\eean
Let us write $Q_i(x)=A_{i,k}(x) B_{i,k}(x) C_{i,k}(x)$ with
\bean
A_{i,k}(x) &=& \prod_{j=k+2}^{^{\ m}_{\longleftarrow}}
R_{i,n+j}(x-2\gamma h^{(i+1,n)}-2\gamma h^{(n+j+1,n+m)}), \\
B_{i,k}(x) &=& R_{i,n+k+1}(x-2\gamma h^{(i+1,n)}-2\gamma h^{(n+k+2,n+m)})
             R_{i,n+k}(x-2\gamma h^{(i+1,n)}-2\gamma h^{(n+k+1,n+m)}), \\
C_{i,k}(x) &=&  \prod_{j=1}^{^{k-1}_{\longleftarrow}}
R_{i,n+j}(x-2\gamma h^{(i+1,n)}-2\gamma h^{(n+j+1,n+m)}). \\
\eean
Since $R_{i,n+k+2} \ldots R_{i,n+m}(x)$ and
$\hat{R}_{n+k,n+k+1}(x)$ act on different spaces and since $[h^{(i)}+h^{(j)},
R_{ij}]=0$, we have
\bean
\hat{R}_{n+k,n+k+1}(x-2\gamma h^{(i,n)}-2\gamma h^{(n+k+2,n+m)})
A_{i,k}(x) && \\
&& \hspace{-8cm} = A_{i,k}(x) \hat{R}_{n+k,n+k+1}(x-2\gamma h^{(i,n)}
-2\gamma h^{(n+k+2,n+m)}).
\eean
Using then the Yang-Baxter equation
$$ \hat{R}_{23}(x-2\gamma h^{(1)}) R_{13}(x) R_{12}(x-2\gamma h^{(3)})
= R_{13}(x) R_{12}(x-2\gamma h^{(3)}) \hat{R}_{23}(x) $$
we see that
\bean
\hat{R}_{n+k,n+k+1}(x-2\gamma h^{(i,n)}-2\gamma h^{(n+k+2,n+m)})
B_{i,k}(x) && \\
&& \hspace{-8cm} = B_{i,k}(x) \hat{R}_{n+k,n+k+1}(x-2\gamma h^{(i+1,n)}
-2\gamma h^{(n+k+2,n+m)}).
\eean
For the same reasons as for $A_{i,k}$,
\bean
\hat{R}_{n+k,n+k+1}(x-2\gamma h^{(i+1,n)}-2\gamma h^{(n+k+2,n+m)})
C_{i,k}(x) && \\
&& \hspace{-8cm} = C_{i,k}(x) \hat{R}_{n+k,n+k+1}(x-2\gamma h^{(i+1,n)}
-2\gamma h^{(n+k+2,n+m)})
\eean
which ends the proof.
\blsquare

Relation $(\ast \ast)$ is proved in a similar way : writing
$Q_n^{-1} \ldots Q_1^{-1} = S_{n+1}^{-1} \ldots S_{n+m}^{-1}$
where
$$ S_j^{-1}(x)=\prod_{k=1}^{^{\ n}_{\longleftarrow}} R_{k,j}^{-1}
(x-2\gamma h^{(k+1,n)}-2\gamma h^{(j+1,n+m)}) $$
and introducing
$$ T'_j(x)={\cal I}^{(n)}(x-2\gamma h^{(j+1,n+m)}) $$
we have, similarly to lemma \ref{lem3},
$$ S_j^{-1}(x) T'_j(x) = T'_{j-1}(x) S_j^{-1}(x). $$

This ends the proof of Theorem (4.1). \blsquare

\section{Examples of commuting Hamiltonians}

We give two examples of application of the above theorem.
In the first one we construct a Ruijsenaars type Hamiltonian with scalar
coefficients. The limit $q \to 1$ yields the usual trigonometric Calogero-Moser
Hamiltonian. In the second example, we construct a set of commuting finite
difference operators with matrix coefficients. Their limit $q \to 1$ is related
to
the spin generalization of the Calogero-Moser model.

To avoid problems of handling infinite dimensional representations of
affine Lie algebras, we restrict ourselves to the trigonometric case
where we need to consider only a finite dimensional matrix algebra.
We shall construct here the quantum analogs of the classical quantities
$\tr L^n(\la=+\infty)$.  We thus
apply theorem \ref{theo} with the $R$-matrix $R^+_{GN}(x,q)$ which
is the limit of eq.(\ref{RF-RGN}) when $\la \longrightarrow +\infty.$
 In the spin Calogero-Moser case, we recall that  these commuting
 Hamiltonians are precisely those which are Yangian-invariant
\cite{BeGauHaPa,BiAvBa}.

\subsection{The scalar case.}
As required by the theorem \ref{theo}, we need  representations of the
algebra (\ref{RLL}) admitting a non trivial subspace of zero weights.
We shall first consider the representation of algebra (\ref{RLL})
analogous to the representation by a completely
symmetrized tensor product $N^{\otimes N}$ of the
Lie algebra $\mbox{sl}_N$ .

By comparison of eq.(\ref{RLL}) and eq.(\ref{GNF}) we see that
$\widetilde{L}^{(N)}_{1 \q}=R_{1 \q}$ is a solution of eq.(\ref{RLL}).
$\widetilde{L}^{(N)}_{1 \q}$ is a matrix in an auxiliary space $(1)$
$$ \widetilde{L}^{(N)}_{1 \q} = \sum_{i,j=1}^N e_{ij}^{(1)}
\widetilde{L}^{(N)}_{ij} $$
the elements of which are quantum operators
represented as the following matrices $\widetilde{L}^{(N)}_{ij}$:
\bea
\label{LN1}
\widetilde{L}^{(N)}_{ii} &=& e_{ii} + \sum_{_{\ j \neq i}^{\ j=1}}^N
\frac{\e^{x_{ij}} - q^{-2} \e^{-x_{ij}}}{\e^{x_{ij}} - \e^{-x_{ij}}} e_{jj} \\
\label{LN2}
\widetilde{L}^{(N)}_{ij} &=& -(1-q^{-2}) \frac{\e^{-x_{ij}}}
{\e^{x_{ij}} - \e^{-x_{ij}}} e_{ji} \quad \mbox{for} \  i\neq j.
\eea
Choosing on the auxiliary space
$$ h_i^{(1)}= (e_{ii}-\frac{1}{N} \mbox{Id}), \quad i=1, \ldots, N, $$
the solution (\ref{LN1},\ref{LN2}) satisfies eq.(\ref{hL}) with
\beq
\label{hN}
h_i^{(\q)} = h_i^{(N)} = (e_{ii}-\frac{1}{N} \mbox{Id}), \quad i=1, \ldots,
N.
\eeq
This is the analog of the vector representation $N$ of $\mbox{sl}_N$.
Next, following \cite{Fel1} one can construct the tensor product of
 $N$ such representations
 \begin{eqnarray}
 \widetilde{L}^{(N^{\otimes N})}=\prod_{j=2}^{N+1}
  \widetilde{L}_{1j} (x -\gamma \sum_{1<i<j}h^{(j)} + \gamma\sum_{j<i\leq
N+1}h^{(j)})
  \nonumber
  \end{eqnarray}

As in the Lie algebra case it turns out that there is a
unique zero weight vector
$$ |V \rangle =  e_1 \otimes e_2\cdots \otimes e_N + permutations $$
Hence
\begin{eqnarray}
\left( tr L^{(N^{\otimes N})} \right) |V \rangle = \widetilde{{\cal H}}  |V
\rangle
\nonumber
\end{eqnarray}

We find after some calculation
\begin{eqnarray}
\widetilde{{\cal H}}= \sum_{i=1}^N e^{2\gamma p_i} \prod_{j \neq i}
\left( {{q^2 e^{x_{ij}} - q^{-2}e^{-x_{ij}}} \over{ q e^{x_{ij}} -
q^{-1}e^{-x_{ij}}}} \right)
\nonumber
\end{eqnarray}
One can perform a similarity transformation
\begin{eqnarray}
{\cal H}=f(x) \widetilde{{\cal H}} {1\over f(x)}; \quad
f(x)= \prod_{k<l} {{e^{x_{kl}}-e^{-x_{kl}}}\over
{ (qe^{x_{kl}}-q^{-1}e^{-x_{kl}})(q^{-1}e^{x_{kl}}-qe^{-x_{kl}})}}
\nonumber
\end{eqnarray}
to get
\begin{eqnarray}
{\cal H}=  \sum_{i=1}^N e^{2\gamma p_i} \prod_{j \neq i}
\left( {{(q e^{x_{ij}} - q^{-1}e^{-x_{ij}})(q^{-1} e^{x_{ij}} - qe^{-x_{ij}})}
\over{ ( e^{x_{ij}} - e^{-x_{ij})^2}}} \right)
\nonumber
\end{eqnarray}
in this form the limit $q=e^{-2\gamma} \to 1$ becomes simple. We find
\begin{eqnarray}
{\cal H} = N +2\gamma \sum_{i=1}^N p_i
+ (2\gamma)^2 \left( {1\over 2}\sum_i p_i^2 -
\sum_{i\neq j}{1\over \sinh^2x_{ij}} \right) +O(\gamma^3)
\nonumber
\end{eqnarray}
Thus we recover the usual trigonometric Calogero-Moser hamiltonian.

\subsection{The spin case}
We shall now construct the representation of algebra (\ref{RLL}) analogous to
the representations  $\bar{N}$ of the Lie algebra $\mbox{sl}_N$
and take its tensor product with the representation $N$.
As in the Lie algebra case, the tensor
product will have a structure similar to the standard decomposition
$N \otimes \bar{N} = 1 + ad$ and admit a
subspace of zero weight vectors of dimension $N$. The Hamiltonians we will
construct act in this zero weight subspace.

One can find another solution of eq.(\ref{RLL}), given by (see also
\cite{BoBo})
\bea
\label{LNbar1}
\widetilde{L}^{(\bar{N})}_{ii} &=& e_{ii} + \sum_{_{\ j \neq i}^{\ j=1}}^N
\frac{\e^{x_{ij}} - q^2 \e^{-x_{ij}}}{\e^{x_{ij}} - \e^{-x_{ij}}} e_{jj} \\
\label{LNbar2}
\widetilde{L}^{(\bar{N})}_{ij} &=& (q-q^{-1}) \frac{\e^{-x_{ij}}}
{q \e^{x_{ij}} - q^{-1} \e^{-x_{ij}}} e_{ij} \quad \mbox{for} \  i\neq j.
\eea
Remark that $\widetilde{L}^{(\bar{N})}$ is essentially the transposed
of $\widetilde{L}^{(N)}.$
In this case we have
\beq
\label{hNbar}
h_i^{(\q)} = h_i^{(\bar{N})} = - (e_{ii}-\frac{1}{N} \mbox{Id}),
\quad i=1, \ldots, N.
\eeq
Notice the sign difference between eq.(\ref{hN}) and eq.(\ref{hNbar}).
Following \cite{Fel1} one now constructs the tensor product of the
two representations:
\beq
\label{Ltensor}
\widetilde{L}_{ij}^{(N \otimes \bar{N})}(x) = \sum_{k=1}^N
\widetilde{L}_{ik}^{(N)}(x+\gamma h^{(\bar{N})})
\widetilde{L}_{kj}^{(\bar{N})}(x-\gamma h^{(N)})
\eeq
and
\beq
\label{hNNbar}
h_i^{(\q)} = h_i^{(N \otimes \bar{N})} =  (e_{ii}-\frac{1}{N} \mbox{Id})
\otimes \mbox{Id} - \mbox{Id} \otimes (e_{ii}-\frac{1}{N} \mbox{Id})
\quad \mbox{for \ } i=1, \ldots, N.
\eeq
{From} this last formula, we see that the subspace of zero weight vectors
admits
$\{ E_i = e_i \otimes e_i \}_{i=1 \ldots N}$ as a basis. We introduce the
canonical basis $\{ E_{ij} \}_{i,j=1 \ldots N}$
of matrices acting on this subspace by $E_{ij} E_j = E_i.$

Applying Theorem \ref{theo} to the $L$ operator
$$ L_{1 \q} = \sum_{i,j=1}^N e_{ij}^{(1)} \ \left [ \e^{\gamma p_i}
\widetilde{L}^{(N \otimes \bar{N})}_{ij}(x) \e^{\gamma p_j} \right ] $$
given by formulas
(\ref{LN1},\ref{LN2},\ref{LNbar1},\ref{LNbar2},\ref{Ltensor}),
we find
\bean
I_1 &=& {\cal H}_1 \\
I_2 &=& I_1^2-(1+q^{-2}) {\cal H}_2 \\
I_3 &=& - \frac{1}{1+q^{-2}} I_1^3 + (1+\frac{1}{1+q^{-2}}) I_2 I_1
+ (1+q^{-2}+q^{-4}) {\cal H}_3
\eean
where the operators ${\cal H}_{1, 2, 3}$ are
\bean
{\cal H}_1 &=& \sumi \e^{2 \gamma p_i} [ \mbox{Id}
+ q^2 (1-q^{-2})^2 \sum_{_{\ j \neq i}^{j=1}}^N V_{ji}(x)
(E_{ji}-E_{jj}) ], \\
{\cal H}_2 &=& \sumij \e^{2 \gamma (p_i+p_j)} [ \frac{1}{2} \mbox{Id}
+ (1-q^{-2})^2 \sum_{_{\ k \neq i,j}^{k=1}}^N V_{kji}(x)
(E_{ki}-E_{kk}) ], \\
{\cal H}_3 &=& \sum_{_{\ i \neq j \neq k }^{i,j,k=1}}^N
\e^{2 \gamma (p_i+p_j+p_k)} [\frac{1}{6} \mbox{Id} + \frac{1}{2}
q^{-2} (1-q^{-2})^2 \sum_{_{\ l \neq i,j,k}^{l=1}}^N V_{lkji}(x)
(E_{li}-E_{ll}) ], \\
\eean
and
\bean
V_{ji}(x) &=& \frac{1}{(\e^{x_{ij}}-\e^{-x_{ij}} )^2}, \\
V_{kji}(x) &=& \frac{1}{(\e^{x_{ik}}-\e^{-x_{ik}} )^2}
\frac{(q^2 \e^{x_{ij}}-\e^{-x_{ij}})(q^2 \e^{x_{jk}}-\e^{-x_{jk}})}
{(\e^{x_{ij}}-\e^{-x_{ij}})(\e^{x_{jk}}-\e^{-x_{jk}})}, \\
V_{lkji}(x) &=& \frac{1}{(\e^{x_{il}}-\e^{-x_{il}} )^2}
\frac{(q^2 \e^{x_{ij}}-\e^{-x_{ij}})(q^2 \e^{x_{jl}}-\e^{-x_{jl}})
(q^2 \e^{x_{ik}}-\e^{-x_{ik}})(q^2 \e^{x_{kl}}-\e^{-x_{kl}})}
{(\e^{x_{ij}}-\e^{-x_{ij}})(\e^{x_{jl}}-\e^{-x_{jl}})
(\e^{x_{ik}}-\e^{-x_{ik}})(\e^{x_{kl}}-\e^{-x_{kl}})}.
\eean
Generalizing the preceding formulas, we introduce quantities $\{ {\cal H}_n \}
_{n=1, \ldots, N}$ defined as
\beq
{\cal H}_n = \sum_{_{i_1 \neq i_2 \neq \ldots \neq i_n}^
{\ i_1, \ldots ,i_n=1}}^N e^{2 \gamma \sum_{k=1}^n p_{i_k}}
\left ( \frac{1}{n !}\mbox{Id} + \frac{q^{-2(n-2)}(1-q^{-2})^2}{(n-1) ! }
 \sum_{_{i_0 \neq i_1, \ldots, i_n}^{\quad i_0=1}}^N
V_{i_0 i_n \ldots i_1}(x) (E_{i_0 i_1}-E_{i_0 i_0})
\right )
\eeq
with
\bea
V_{i_0 i_n \ldots i_1}(x) &=& V_{i_0 i_{n-1} \ldots i_1}(x)
\frac{(q^2 \e^{x_{i_1 i_n}}-\e^{-x_{i_1 i_n}})
(q^2 \e^{x_{i_n i_0}}-\e^{-x_{i_n i_0}})}{(\e^{x_{i_1 i_n}}-\e^{-x_{i_1 i_n}})
(\e^{x_{i_n i_0}}-\e^{-x_{i_n i_0}})} \\
V_{i_0 i_1}(x) &=& \frac{1}{(\e^{x_{i_0 i_1}}-\e^{-x_{i_0 i_1}} )^2}.
\eea
We have checked directly, up to 5 particles, that the $\{ {\cal H}_n \}_{n=1,
\ldots, N}$
form a set of commuting operators.

The occurence of the matrices $E_{ji}-E_{jj}$ immediately shows that the
vector
$$ |s \rangle = \sumi e_i \otimes e_i $$
is left invariant by all the Hamiltonians ${\cal H}_n.$ Their restriction to
this one dimensional subspace is the abelian algebra of the symmetric
polynomials in $\e^{2 \gamma p_i}.$

To recover the usual Calogero-Moser Hamiltonian we have to consider the
expansion around $\gamma = 0$  of the above Hamiltonians. To order
$\gamma ^2$ we find
\beq
{\cal H}_n = \CC_N^n \, \mbox{Id} + 2 \gamma \, \CC_{N-1}^{n-1} \, {\cal H}_1^
{CM} + 2 \gamma^2 \left [ \CC_{N-2}^{n-2} \left (
{\cal H}_1^{CM} \right ) ^2 + \CC_{N-2}^{n-1} \,
{\cal H}_2^{CM} \right ] + \mbox{O}(\gamma^3)
\eeq
where $\CC_{N}^{n}$ are the usual binomial coefficients and
\bea
  {\cal H}_1^{CM} &=& \sumi \frac{\partial}{\partial x_i} \\
  {\cal H}_2^{CM} &=& - \frac{1}{2} \sumi \frac{\partial^2}
{\partial x_i^2} - \sum_{_{\ i < j}^{i,j=1}}^N \frac{1}{\sinh^2(x_{ij})}
( E_{ij} + E_{ji} - E_{ii} - E_{jj} ).
\eea
The matrices $ E_{ij} + E_{ji} - E_{ii} - E_{jj} $ admit a simple
interpretation in term of the ``spin operator'' $h_{ij}$ in the tensor
product representation $N \otimes \bar{N}$
$$ h_{ij} = e_{ij} \otimes \mbox{Id} - \mbox{Id} \otimes e_{ji}. $$
Indeed we have
$$ \left. h_{ij} h_{ji} \right \vert_{\, zero~weight} =
E_{ii} + E_{jj} - E_{ij} - E_{ji}, $$
hence in this representation we do recover the spin Calogero-Moser Hamiltonian
\beq
{\cal H}_2^{CM} = - \frac{1}{2} \sumi \frac{\partial^2}{\partial
x_i^2} + \sumij h_{ij} h_{ji} \frac{1}{\sinh^2(x_{ij})}
\eeq
as the first non-trivial order of ${\cal H}_n.$

We would like to stress that the above examples are built out of the simplest
representations of (\ref{RLL}) admitting a non-trivial zero-weight
subspace. More general representations will affect, among other things, the
value of the
coupling constant. As indicated by the $\mbox{sl}_2$ case \cite{Ba1}, the
representation
theory of eq.(\ref{RLL}) is intimately tied to the representation theory of
quantum groups, but the link remains to be fully elucidated.


\bigskip

\bigskip

\begin{thebibliography}{**}

\bibitem{CalMo} F. Calogero, {\it Exactly solvable one-dimensional many-body
problems}, Lett. Nuovo Cimento {\bf 13},~411~(1975). {\it On a functional
equation connected with integrable many-body problems}, Lett. Nuovo
Cimento {\bf 16}, ~77~(1976). \\
J. Moser, {\it Three integrable Hamiltonian systems
connected to isospectral deformations}, Adv. Math. {\bf 16}, ~1~(1976).

\bibitem{Rui} S.N.M. Ruijsenaars, {\it Complete integrability of relativistic
Calogero-Moser systems and elliptic function identities}, Commun. Math. Phys.
{\bf 110}, ~191~(1987). \\
J.F. van Deijen, {\it Families of commuting difference operators}, Ph.D.
Thesis, University of Amsterdam (1994).

\bibitem{GibHe} J. Gibbons, T. Hermsen, {\it A generalisation of the
Calogero-Moser system}, Physica {\bf 11D},~337~(1984). \\
I.M. Krichever, O. Babelon, E. Billey, M. Talon, {\it Spin generalization
of Calogero-Moser system and the matrix KP equation}, hep-th/9411160.

\bibitem{BeGauHaPa} D. Bernard, M. Gaudin, F.D.M. Haldane, V. Pasquier,
{\it Yang-Baxter equation in spin chains with long range interactions},
J. Phys. A: Math. Gen. {\bf 26},~5219~(1993), hep-th/9301084.

\bibitem{KriZa} I. Krichever, A. Zabrodin, {\it Spin generalization of the
Ruijsenaars-Schneider model, non-abelian 2D Toda chain and representations
of Sklyanin algebra}, hep-th/9505039.

\bibitem{OlPer} M.A. Olshanetsky, A.M. Perelomov, {\it Classical
integrable finite-dimensional systems related to Lie algebras}, Phys. Rep.
{\bf 71},~314~(1981). {\it Quantum integrable systems related to Lie
algebras}, Phys. Rep. {\bf 94}, 313 (1983).

\bibitem{HiNe} N. Hitchin, {\it Stable bundles and integrable systems},
Duke Math. Journ. {\bf 54},~91~(1987). \\
N. Nekrasov, {\it Holomorphic bundles and many-body systems}, hep-th/9503157.

\bibitem{GorNe} A. Gorsky, N. Nekrasov, {\it Elliptic Calogero-Moser
system from two-dimensional current algebra}, Nucl. Phys. {\bf B436},
{}~532~(1995),
hep-th/9401021.
{\it Relativistic Calogero-Moser model as gauged WZW theory},
Theor. Math. Phys. {\bf 100},~874~(1994), hep-th/9401017. \\
A. Gorsky, {\it Integrable many body systems in the
field theories}, hep-th/9410228.

\bibitem{Fel1} G. Felder, {\it Conformal field theory and integrable
systems associated to elliptic curves}, hep-th/9407154.

\bibitem{EtKir} P.I. Etingof, A.A. Kirillov, {\it On the affine analogs
of Jack's and Macdonald's polynomials}, hep-th/9403168. {\it MacDonald's
polynomials and
representations of quantum groups}, Math. Res. Lett. {\bf 1}, 279 (1994).

\bibitem{AvJev} J. Avan, A. Jevicki, {\it Classical integrability
and higher symmetries of collective string field theory}, Phys. Lett. {\bf B
266},~35~(1991).

\bibitem{AvTaSkBraSu} J. Avan, M. Talon, {\it Classical $R$-matrix
structure for the Calogero model}, Phys. Lett. {\bf B 303},~33~(1993). \\
E.K. Sklyanin, {\it Dynamical $r$-matrices for the
elliptic Calogero-Moser model}, Alg. and Anal. {\bf 6},~227~(1994)
, hep-th/9308060. \\
H.W. Braden, T. Suzuki, {\it $R$-matrices for elliptic Calogero-Moser
models}, Lett. Math. Phys. {\bf 30},~147~(1994), hep-th/9309033.

\bibitem{BiAvBa} E. Billey, J. Avan, O. Babelon, {\it The $r$-matrix
structure of the Euler-Calogero-Moser model}, Phys. Lett. {\bf A 186},
{}~114~(1994),
hep-th/9312042.
{\it Exact Yangian symmetry in the classical Euler-Calogero-Moser
model}, Phys. Lett. {\bf A 188},~263~(1994), hep-th/9401117.

\bibitem{BaBe} O. Babelon, D. Bernard, {\it The sine-Gordon solitons as
an $N$-body problem}, Phys. Lett. {\bf B 317},~363~(1993).

\bibitem{AvRol} J. Avan, G. Rollet, {\it The classical $r$-matrix for
the relativistic Ruijsenaars--Schneider system}, Preprint Brown HET 1014
(1995).

\bibitem{Kri} I. Krichever, {\it Elliptic solutions of the Kadomtsev-
Petviashvili equation and integrable systems of particles}, Func. Anal.
Appl. {\bf 14},~282~(1980).

\bibitem{BaVi} O. Babelon, C.M. Viallet, {\it Hamiltonian structures and
Lax equations}, Phys. Lett. {\bf B 237},~411~(1990).

\bibitem{KnZaBerFelWi} V.G. Knizhnik, A.B. Zamolodchikov, {\it Current algebra
and Wess-Zumino model in two dimensions}, Nucl. Phys. {\bf B 247},
 ~83~(1984). \\
D. Bernard, {\it On the Wess-Zumino-Witten models on the torus},
Nucl. Phys. {\bf B 303},~77~(1988). {\it On the Wess-Zumino-Witten models on
Riemann surfaces}, Nucl. Phys. {\bf B 309},~145~(1988). \\
G. Felder, C. Wieczerkowski, {\it Conformal field
theory on elliptic curves and Knizhnik-Zamolodchikov-Bernard equation},
hep-th/9411004.

\bibitem{LezSav} A.N. Leznov, M.V. Saveliev, {\it Representation of zero-
curvature for the system of non-linear partial differential equations
$x_{\alpha, z \bar{z}} = \exp(h x)_{\alpha}$ and its
integrability}, Lett. Math. Phys. {\bf 3},~489~(1979).

\bibitem{BaBoToBaDaFe} O. Babelon, L. Bonora, F. Toppan, {\it Exchange
algebra and the Drinfeld-Sokolov theorem}, Commun. Math. Phys. {\bf 140},
{}~93~(1991). \\
J. Balog, L. Dabrowski, L. Feh\'er, {\it Classical
$r$-matrix and exchange algebra in WZNW and Toda theories}, Phys. Lett. {\bf B
244},~227~(1990).

\bibitem{GeNe} J.L. Gervais, A. Neveu, {\it Novel triangle relation
and absence of tachyons in Liouville string field theory}, Nucl. Phys. {\bf B
238},~125~(1984).

\bibitem{BiGe} A. Bilal, J.L. Gervais, {\it Systematic construction of
conformal theories with higher spin Virasoro symmetries}, Nucl. Phys. {\bf B
318}, ~579~(1989).

\bibitem{BoBo} L. Bonora, V. Bonservizi, {\it Quantum sl$_n$ Toda
field theories}, Nucl. Phys. {\bf B 390}, ~205~(1993).

\bibitem{Fadd} L.D. Faddeev, {\it Integrable models in $1+1$ dimensional
quantum field theory}, Recent Advances in Field Theory and Statistical
Mechanics (Les Houches Summer School 1982), ed. by J.B. Zuber and R. Stora,
Amsterdam, North Holland (1984).

\bibitem{Ba1} O. Babelon, {\it Universal exchange algebra for Bloch waves
and Liouville theory}, Commun. Math. Phys. {\bf 139}, ~619~(1991).

\bibitem{Ba2} O. Babelon, {\it Extended conformal algebra and the
Yang-Baxter equation}, Phys. Lett. {\bf B 215},~523~(1988).

\bibitem{CrGer} E. Cremmer, J.L. Gervais, {\it The quantum group structure
associated with non-linearly extended Virasoro algebras}, Commun. Math.
Phys. {\bf 134},~619~(1990).

\bibitem{FoIoJiKeMiYa} O. Foda, K. Iohara, M. Jimbo, R. Kedem, T. Miwa,
H. Yan, {\it An elliptic quantum algebra for $\hat{sl_2}$},
Lett. Math. Phys. {\bf 32},~259~(1994), hep-th/9403094.

\bibitem{Mail} J.M Maillet, {\it Lax equations and quantum groups},
Phys. Lett. {\bf B 245}, ~480~(1990).

\end{thebibliography}
\end{document}